\documentclass[onecolumn, 12pt]{IEEEtran}
\usepackage{amsmath,amsfonts,amssymb,enumerate,cite}
\usepackage[final]{lscape,graphicx,epsfig}
\usepackage{cite}
\usepackage{setspace}

\DeclareMathOperator*{\argmin}{argmin}

\newcommand{\lp}{\ensuremath{\theta}}
\newcommand{\scpar}{\ensuremath{\sigma}}
\newtheorem{thm}{Theorem}

\title{Distributed SNR Estimation using Constant Modulus Signaling over Gaussian Multiple-Access Channels}
\author{Mahesh K. Banavar, {\em Student Member, IEEE}, Cihan Tepedelenlio\u{g}lu, {\em Member,
IEEE} and Andreas
Spanias, {\em Fellow, IEEE} \thanks{The work in this paper is
supported by the SenSIP Center, Arizona State University. The authors are with the SenSIP
Center, School of ECEE,
Fulton Schools of Engineering, Arizona State University. (Email: \{maheshkb, cihan, spanias\}@asu.edu). }}

\date{}

\begin{document}

\maketitle
\bibliographystyle{IEEEtran}

\begin{abstract}
A sensor network is used for distributed joint mean and variance estimation, in a single time snapshot. Sensors observe a signal embedded in noise, which are phase modulated using a constant-modulus scheme and transmitted over a Gaussian multiple-access channel to a fusion center, where the mean and variance are estimated jointly, using an asymptotically minimum-variance estimator, which is shown to decouple into simple individual estimators of the mean and the variance. The constant-modulus phase modulation scheme ensures a fixed transmit power, robust estimation across several sensing noise distributions, as well as an SNR estimate that requires a single set of transmissions from the sensors to the fusion center, unlike the amplify-and-forward approach. The performance of the estimators of the mean and variance are evaluated in terms of asymptotic variance, which is used to evaluate the performance of the SNR estimator in the case of Gaussian, Laplace and Cauchy sensing noise distributions. For each sensing noise distribution, the optimal phase transmission parameters are also determined. The asymptotic relative efficiency of the mean and variance estimators is evaluated. It is shown that among the noise distributions considered, the estimators are asymptotically efficient only when the noise distribution is Gaussian. Simulation results corroborate analytical results. 
\end{abstract}


\section{Introduction}
\label{sec:introduction}

SNR estimation of a signal embedded in noise finds applications in diverse areas in signal processing and communications, such as signal strength estimation for cognitive radio, in diversity combining and in bit-synchronization applications. The SNR estimate can be obtained by combining the estimates of the mean and the variance. In addition to being used to form the SNR estimate, the variance can be used to estimate the quality and the variability of the sensor measurements, since the quality of the estimates of the mean and variance in turn depend on the true value of the variance \cite{Slijepcevic2003, Ashraf2009}. 

In centralized estimation problems, the samples of the signals embedded in noise are directly available to the estimator \cite{Koutrouvelis1980, Koutrouvelis1981, Feuerverger1981, Koutrouvelis1982, Eubank1982, Pauluzzi2000}. Centralized schemes for SNR estimation of signals embedded in Gaussian noise are considered in \cite{KayEst, Pauluzzi2000, vanTrees1968, Kerr1966, Benedict1967, Matzner1993}. In the case of non-Gaussian noise, the mean and variance are estimated separately, and then combined to form the SNR estimate, as discussed in \cite{Koutrouvelis1980, Koutrouvelis1981, Feuerverger1981, Koutrouvelis1982, Eubank1982}. In \cite{Koutrouvelis1980, Koutrouvelis1981, Feuerverger1981, Koutrouvelis1982}, the mean and the variance are separately estimated from the characteristic function of the signal embedded in noise. 

In contrast to the centralized regime outlined above, in the case of distributed estimation, the observations are not directly available at the estimator, but have to be transmitted to a fusion center (FC) for estimation. In \cite{Kapnadak2008, Senel2008}, the authors consider a digital transmission scheme where the sensors quantize their observations and then transmit them to the FC. Analog transmissions can also be used between the sensors and the FC. The most commonly used type of analog transmission is the amplify-and-forward scheme, \cite{gastparb2003, maheshjrnl}, where the instantaneous transmit power at the sensors depends on the individual sensing noise realizations, and can be arbitrarily high in the typical case where the sensing noise has infinite support. Furthermore, with amplify-and-forward transmissions, the signal received at the FC approximates the sample mean of the sensor observations. Since with amplify-and-forward, a linear function of the sensed data is transmitted over additive Gaussian multiple-access channels, only information about the first moment is available at the FC. In order to estimate other (higher-order) moments with amplify-and-forward transmissions from the sensors to the FC, each sensor will have to transmit different powers of the data across multiple time snapshots, which is not desirable in delay and bandwidth sensitive applications. When estimation is attempted at the sensors before transmission \cite{Senel2008}, multiple observations are required to accumulate at the sensors before the estimation, making estimation in a single time snapshot difficult. In contrast, transmitting non-linear functions of the observation makes it possible to estimate second moments within a single time-slot, making such a scheme preferable for delay-sensitive applications \cite{Goldenbaum2009, Goldenbaum2010}. 

\begin{figure}[tb]
\begin{minipage}[b]{1.0\linewidth}
  \centering
  \centerline{\epsfig{figure=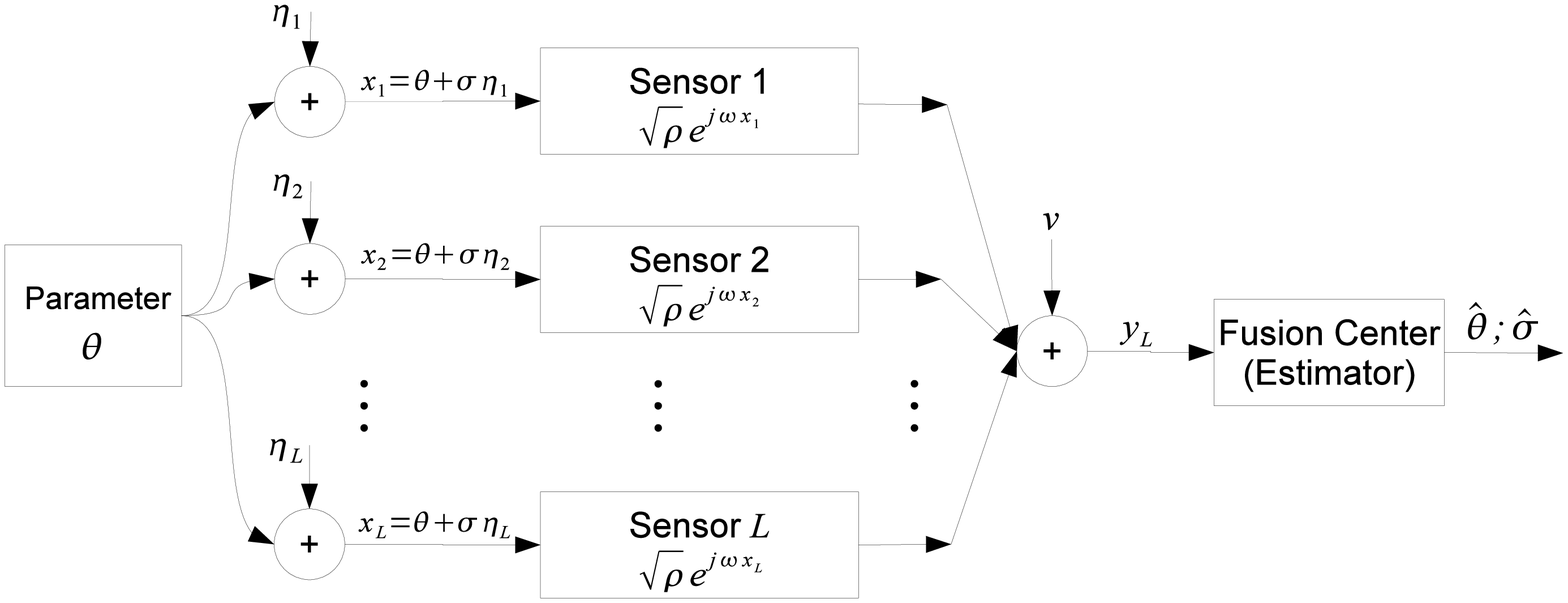,width=11cm,height=5cm}}
\end{minipage}
\caption{System model: Wireless sensor network with constant modulus transmissions from the sensors.} \label{Fig:problem_setup}
\end{figure}

In this paper, we consider the joint estimation of the mean and variance of a signal embedded in noise in a distributed fashion, for the first time in the literature. Sensors are exposed to a signal in (not necessarily Gaussian) noise as seen in Figure \ref{Fig:problem_setup}. The sensors phase modulate the observations using a constant-modulus scheme and transmit these signals to a fusion center (FC) over a Gaussian multiple-access channel \cite[pp. 378]{Cover91}. These analog transmissions are appropriately pulse-shaped to consume finite bandwidth. Similar to \cite{robust_est_cihan_adarsh_2010}, the constant-modulus nature of the transmissions ensures a fixed instantaneous transmit power at the sensors, irrespective of the sensing noise realizations. Due to the additive nature of the multiple-access channel, the signals transmitted from the sensors superimpose at the FC, and approximate the characteristic function of the sensed data. This enables robust estimation of not only the mean but also the variance, unlike \cite{robust_est_cihan_adarsh_2010}. 
These estimates of the mean and variance are used for constructing the SNR estimate, and therefore, the phase modulation scheme allows for the estimation of the mean, variance and SNR with a single set of transmissions from the sensors to the FC. 
The asymptotic relative efficiency of each of the estimators is calculated for different sensing noise distributions. It is shown that among the sensing noise distributions considered (Gaussian, Laplace and Cauchy), the estimators are asymptotically efficient only if the sensing noise distribution is Gaussian. 
In cases where the moments of the sensed data do not exist, the more general quantities, location parameter and scale parameter are defined and used. 

The rest of this paper is organized as follows. The system model is introduced in Section \ref{sec:model}. In Section \ref{sec:tpc}, it is assumed that there is a fixed total power available across all sensors. Minimum-variance estimators for the mean and variance are presented, along with optimal values for the phase modulation parameter for different sensing noise distributions. In Section \ref{sec:pspc}, the estimators are revisited assuming a fixed-power budget at each sensor. The asymptotic relative efficiency of each of the estimators is also found under this per-sensor power constraint in Section \ref{sec:are}. Simulation results are presented in Section \ref{sec:simulations}, and concluding remarks in Section \ref{sec:conclusions}. 

\section{System Model}
\label{sec:model}

A sensor network, illustrated in Figure \ref{Fig:problem_setup}, consisting of $L$ sensors observe a deterministic parameter, $\lp$, in noise. The value, $x_{l}$, observed at the
$l^{th}$ sensor is
\begin{equation}
x_{l} = \lp + \scpar \eta_l \label{eqn:obs_eqn}
\end{equation}
for $l = 1,...,L$, where $\lp$ is a deterministic, real-valued, unknown parameter in a bounded interval, $(0, \lp_{R}]$, of known length, $\lp_{R}<\infty$, and $\eta_{l}$ are independent and identically distributed (iid) real-valued random variables drawn from a distribution symmetric about zero, and $\scpar > 0$ is a scale parameter, which is proportional to the standard deviation when the standard deviation of $\eta_{l}$ exists. When the mean of $\eta_{l}$ exists, $E[x_{l}] = \lp$. Otherwise, we will refer to $\lp$ as a {\em location parameter} of $x_{l}$. 
The sensing SNR is defined as $\gamma \mathop{:=} \lp^{2}/\scpar^{2}$. Due to practical constraints on the peak transmit power, we consider a scheme where the $l^{th}$ sensor transmits its measurement, $x_{l}$, using a constant modulus base-band equivalent signal, $\sqrt{\rho} e^{j \omega x_{l}}$, with power $\rho$, over a Gaussian multiple-access channel so that the received signal at the fusion center is given by
\begin{equation}
y_{L} = \sqrt{\rho} \sum_{l=1}^{L} e^{j \omega x_{l}} + \nu,
\label{eqn:yn}
\end{equation}
where $\omega\in(0,2\pi /\lp_{R}]$, is a
design parameter to be optimized, and $\nu \sim \mathcal{CN}(0,\sigma_{\nu}^{2})$ is the channel noise independent of $\{\eta_{l}\}_{l=1}^{L}$. Note that the restriction $\omega\in(0,2\pi /\lp_{R}]$ is necessary even in the absence of sensing and channel noise to uniquely determine $\lp$ from $y_{L}$. 

As seen in (\ref{eqn:yn}), all sensors transmit using the same value of $\omega$. The transmissions are to be appropriately pulse-shaped and phase modulated to consume finite bandwidth. The transmission power at each sensor is the same and is given by $\rho$. Two cases of power constraint are considered in this paper. In the first case, a {\em total power constraint}, $P$, is considered, where $\rho = P/L$. Irrespective of the number of sensors in the system, the total transmit power from all the sensors in the system remains $P$. The other transmission scheme is a \textit{per-sensor power constraint}, where $\rho = P$. In this regime, increasing the number of sensors increases the total transmit power, and as $L\to\infty$, the channel noise becomes negligible compared to the transmit power. 

\section{Total Power Constraint}
\label{sec:tpc}

Under the total power constraint on the sensor transmissions, we derive asymptotically optimal estimators in this section. The asymptotic variance of the estimators are also derived, and the value of $\omega$ that minimizes the asymptotic variance is computed. 

In the total power constraint regime, each sensor transmits with a power of $\rho = P/L$. The normalized signal at the FC that the estimator acts on is given by
\begin{equation}
z_{L} \mathop{:=} \frac{y_{L}}{\sqrt{L}} = \sqrt{P} \frac{1}{L} \sum_{i=1}^{L} e^{j \omega x_{i}} + \frac{\nu}{\sqrt{L}}.
\label{eqn:zl_defn}
\end{equation}
Asymptotically, as $L\to\infty$,
\begin{align}
\bar{z} &\mathop{:=} \lim_{L\to\infty} z_{L} = \sqrt{P} \lim_{L\to\infty} \frac{1}{L} \sum_{i=1}^{L} e^{j \omega x_{i}} \nonumber \\
& = \sqrt{P} e^{j \omega \lp} \varphi_{\eta} (\scpar \omega),
\label{eqn:z_defn}
\end{align}
with probability one, where $\varphi_{\eta} (\scpar \omega) = E \left[ e^{j \omega \scpar \eta_{i}} \right]$ is the characteristic function of $\eta_{i}$. The characteristic function of the sensing noise is real-valued, since the distribution of $\eta_{i}$ is symmetric about the median. Also define $\mathbf{z}_{L} \mathop{:=} [z_{L}^{R} \text{  } z_{L}^{I}]^{T}$ where $z_{L}^{R}$ and $z_{L}^{I}$ are the real and imaginary parts of the random variable, $z_{L}$, respectively. 
The vector $\mathbf{z}_{L}$ converges for large $L$ to $\bar{\mathbf{z}} = [\bar{z}^{R} \text{  } \bar{z}^{I}]^{T}$, where $\bar{z}^{R} = \lim_{L\to\infty} z_{L}^{R} = \text{Re} \{\bar{z}\} = \sqrt{P} \cos (\omega \lp) \varphi_{\eta} (\scpar \omega)$ and $\bar{z}^{I} = \lim_{L\to\infty} z_{L}^{I} = \text{Im} \{\bar{z}\} = \sqrt{P} \sin (\omega \lp) \varphi_{\eta} (\scpar \omega)$. 
Due to the central limit theorem, this convergence takes place in such a way that
$\lim_{L\to\infty} \sqrt{L} (\mathbf{z}_{L} - \bar{\mathbf{z}})$
is a $2\times 1$ Gaussian random vector with zero mean and a $2\times 2$ covariance matrix $\boldsymbol{\Sigma}$ with elements
\begin{align}
\Sigma_{11} &= P \left[ v_{c} \cos^{2}(\omega \lp) + v_{s} \sin^{2}(\omega \lp) \right] + \frac{1}{2} \sigma_{\nu}^{2} \nonumber\\
\Sigma_{22} &= P \left[ v_{s} \cos^{2}(\omega \lp) + v_{c} \sin^{2}(\omega \lp) \right] + \frac{1}{2} \sigma_{\nu}^{2} \nonumber\\
\Sigma_{12} &= \Sigma_{21} = P (v_{c} - v_{s})
\sin(\omega \lp) \cos(\omega \lp),
\label{eqn:sigma_defn}
\end{align}
where the parameters $v_{c} \mathop{:=} \text{var} [\cos (\omega \eta_{l})] = 1/2 +
\varphi_{\eta}(2 \scpar \omega) /2 - \varphi_{\eta}^{2}(\scpar \omega)$ and $v_{s} \mathop{:=}
\text{var}[\sin (\omega \eta_{l})] = 1/2 - \varphi_{\eta}(2 \scpar \omega)/2$.

\subsection{The Asymptotically Minimum Variance Estimator}
\label{ssec:estimator}

From $\mathbf{z}_{L}$ obtained at the FC, the values of $\lp$ and $\scpar$ are estimated. The estimator for $[\lp \text{  } \scpar]^{T}$ which yields the minimum variance is given by \cite[(3.6.2), pp. 82]{porat1994}
\begin{equation}
\left[ \hat{\lp}^{\rm opt}\quad \hat{\scpar}^{\rm opt} \right]^{T} = \argmin_{\lp, \scpar} [\mathbf{z}_{L} -
\bar{\mathbf{z}}]^{T} \boldsymbol{\Sigma}^{-1}
[\mathbf{z}_{L} - \bar{\mathbf{z}}],
\label{eqn:est_lp_scpar}
\end{equation}
where $z_{L}$ represents the normalized received data, and the right-hand-side of (\ref{eqn:est_lp_scpar}) depends on $\lp$ and $\scpar$ through both $\boldsymbol{\Sigma}$ and $\bar{\mathbf{z}}$. Intuitively, if the central limit theorem is invoked on $z_{L}$ in (\ref{eqn:zl_defn}), the ML estimator of $[ \hat{\lp} \quad \hat{\scpar} ]^{T}$ is the same as the estimator in (\ref{eqn:est_lp_scpar}). Interestingly, it is possible to express the asymptotic covariance of the estimator in (\ref{eqn:est_lp_scpar}), without having to express (\ref{eqn:est_lp_scpar}) in closed form. The asymptotic covariance of the asymptotically minimum variance estimator is given by $[ \mathbf{J}^{T} \boldsymbol{\Sigma}^{-1} \mathbf{J} ]^{-1}$ \cite[Lemma 3.1]{porat1994},
where $\mathbf{J}$ is the Jacobian matrix of $\bar{\mathbf{z}}$ with respect to $\lp$ and $\scpar$ and is given by
\begin{equation}
\mathbf{J} = \omega \sqrt{P} \left[ \begin{array}{c c}
- \sin(\omega \lp) \varphi_{\eta} (\omega \scpar) & \cos(\omega \lp) \frac{\partial \varphi_{\eta} (\omega \scpar)}{\partial \scpar} \\
\cos(\omega \lp) \varphi_{\eta} (\omega \scpar) & \sin(\omega \lp) \frac{\partial \varphi_{\eta} (\omega \scpar)}{\partial \scpar}
\end{array} \right]. 
\label{eqn:jacobian}
\end{equation}
After a straightforward calculation, the asymptotic covariance matrix is seen to be diagonal with the elements given by the asymptotic variances of $\hat{\lp}^{\rm opt}$ and $\hat{\scpar}^{\rm opt}$ respectively as
\begin{align}
\text{AsV}_{\hat{\lp}^{\rm opt}} (\omega) &= \frac{P + \sigma_{\nu}^{2} -P \varphi_{\eta} (2\scpar \omega)}{2 P \omega^{2} \varphi_{\eta}^{2} (\scpar \omega)} \label{eqn:asv_lp_tpc} \\
\text{AsV}_{\hat{\scpar}^{\rm opt}} (\omega) &= \frac{P + \sigma_{\nu}^{2} -2P \varphi_{\eta}^{2} (\scpar \omega) + P \varphi_{\eta} (2 \scpar \omega)}{2 P \left[ \frac{\partial \varphi_{\eta}(\scpar \omega)}{\partial \scpar} \right]^{2}}. \label{eqn:asv_scpar_tpc} 
\end{align}

\subsection{Simplified Estimator}
\label{ssec:alt_est}

From the structure of $z_{L}$ in (\ref{eqn:z_defn}) and the characteristic function, $\varphi_{\eta} (\sigma \omega)$, alternative estimators for $\lp$ and $\scpar$ can be constructed. Separating the signal into its magnitude and phase components,
\begin{align}
|z_{L}| &= \sqrt{P} \varphi_{\eta}(\scpar \omega), 
\label{eqn:z_abs} \\
\angle z_{L} &= \omega \lp,
\label{eqn:z_arg}
\end{align}
where (\ref{eqn:z_abs}) depends on $\scpar$ and not $\lp$, whereas (\ref{eqn:z_arg}) depends on $\lp$ and not $\scpar$, and can be used to construct low-complexity estimators. The simple estimates, $\hat{\lp}^{\rm sim}$ and $\hat{\scpar}^{\rm sim}$, are the solutions to (\ref{eqn:z_arg}) and (\ref{eqn:z_abs}), respectively. In what follows, the relationship between these estimators and the minimum-variance joint estimator in (\ref{eqn:est_lp_scpar}) is established. 

\begin{thm}
\label{thm:compare_estimators}
When $|z_{L}|\leq \sqrt{P}$, the estimates $\hat{\lp}$ and $\hat{\scpar}$ that solve (\ref{eqn:z_arg}) and (\ref{eqn:z_abs}), respectively, are those that minimize (\ref{eqn:est_lp_scpar}), that is,
\begin{align}
\hat{\lp}^{\rm opt} &= \hat{\lp}^{\rm sim} \label{eqn:thm_lp eq}\\
\hat{\scpar}^{\rm opt} &= \hat{\scpar}^{\rm sim} \label{eqn:thm_scpar eq}. 
\end{align}
\end{thm}

\begin{IEEEproof}
The proof is presented in Appendix \ref{app:proof_thm_compare_estimators}. 
\end{IEEEproof}

Since $z_{L}$ is asymptotically given by $\bar{z} = \sqrt{P} e^{j \omega \lp} \varphi_{\eta} (\scpar \omega)$ in (\ref{eqn:z_defn}), and because $\varphi_{\eta} (\scpar \omega) \leq 1$, the condition of Theorem \ref{thm:compare_estimators}, $|z_{L}| \leq \sqrt{P}$, is satisfied almost surely for sufficiently large $L$. Therefore, the estimators in (\ref{eqn:thm_lp eq}) and (\ref{eqn:thm_scpar eq}) are identical in the asymptotic regime, and they will be denoted by $\hat{\lp}$ and $\hat{\scpar}$. Their performance will also be the same, denoted by $\text{AsV}_{\hat{\lp}} (\omega)$ and $\text{AsV}_{\hat{\scpar}} (\omega)$ and given by (\ref{eqn:asv_lp_tpc}) and (\ref{eqn:asv_scpar_tpc}), respectively. Recall that (\ref{eqn:asv_lp_tpc}) and (\ref{eqn:asv_scpar_tpc}) are in the diagonal of the asymptotic covariance matrix, indicating that the estimate of the location parameter and the estimate of the scale parameter are asymptotically independent. 

From the scale parameter and the location parameter of $x_{l}$, the sensing SNR can be estimated as 
\begin{equation}
\hat{\gamma} = \frac{\hat{\lp}^{2}}{\hat{\scpar}^{2}}. 
\label{eqn:SNR_est_defn}
\end{equation} 
The estimator of $\gamma$ in (\ref{eqn:SNR_est_defn}) is constructed using the optimal estimators of $\theta$ and $\sigma$. Since the estimates of $\lp$ and $\scpar$ are asymptotically ML, and the function in (\ref{eqn:SNR_est_defn}) is one-to-one, it can be verified using the \textit{invariance property} of the MLE \cite[Thm. 7.2]{KayEst} that $\hat{\gamma}$ is also an asymptotically ML estimate of $\gamma$. 

From the asymptotic variances of $\hat{\lp}$ and $\hat{\scpar}$ and using \cite[pp. 185]{KayEst}, the estimate of $\gamma$ in (\ref{eqn:SNR_est_defn}) has an asymptotic variance given by
\begin{equation}
\text{AsV}_{\hat{\gamma}} (\omega) = \left( \frac{\partial \gamma}{\partial \lp} \right)^{2} \text{AsV}_{\hat{\lp}} (\omega) + \left( \frac{\partial \gamma}{\partial \scpar} \right)^{2}  \text{AsV}_{\hat{\scpar}} (\omega) = \frac{4 \gamma} {\sigma^{2}} \left[ \text{AsV}_{\hat{\lp}} (\omega) + \gamma \text{AsV}_{\hat{\scpar}} (\omega) \right],
\label{eqn:SNR_asv}
\end{equation}
where $\gamma \mathop{:=} \lp^{2}/\scpar^{2}$. 

\subsection{Optimization of $\omega$}
\label{ssec:opt_omega}

Ideally, the sensors should use the value of $\omega$ that minimizes the expressions in (\ref{eqn:asv_lp_tpc}), (\ref{eqn:asv_scpar_tpc}) and (\ref{eqn:SNR_asv}). For many sensing distributions, it will be tractable to minimize $\text{AsV}_{\hat{\lp}} (\omega)$ and $\text{AsV}_{\hat{\scpar}} (\omega)$ with respect to $\omega$. However, $\text{AsV}_{\hat{\gamma}} (\omega)$ in (\ref{eqn:SNR_asv}) will be more involved. We are therefore motivated to relate the minimizer, $\omega_{\gamma}^{*}$, of (\ref{eqn:SNR_asv}) with $\omega_{\lp}^{*}$ and $\omega_{\scpar}^{*}$, which minimize (\ref{eqn:asv_lp_tpc}) and (\ref{eqn:asv_scpar_tpc}), respectively. Specifically, we will show that $\omega_{\gamma}^{*}$ lies between $\omega_{\lp}^{*}$ and $\omega_{\scpar}^{*}$. To do this, we will exploit the fact that for many sensing noise distributions, the expressions for $\text{AsV}_{\hat{\lp}} (\omega)$ and $\text{AsV}_{\hat{\scpar}} (\omega)$ are quasi-convex in $\omega$, and differentiable. A univariate function, $f(\omega)$ on $\omega \in (0,2\pi/\theta_{R}]$, is quasi-convex if it satisfies any one of the following conditions \cite[pp. 99]{boyd2004}: 
\begin{enumerate}
 \item[(c1)] $f(\omega)$ is monotonic, i.e., $f(\omega)$ is non-decreasing or $f(\omega)$ is non-increasing
 \item[(c2)] $f(\omega)$ has a global minimum at $\omega^{*}$ such that for $\omega \leq \omega^{*}$, $f(\omega)$ is non-increasing and for $\omega \geq \omega^{*}$, $f(\omega)$ is non-decreasing.
\end{enumerate}

For settings where $\text{AsV}_{\hat{\lp}} (\omega)$ and $\text{AsV}_{\hat{\scpar}} (\omega)$ are quasi-convex in $\omega$, the following theorem provides a relationship between $\omega_{\lp}^{*}$, $\omega_{\scpar}^{*}$ and $\omega_{\gamma}^{*}$. We will see later in this section that when the sensing noise is Gaussian, Laplace or Cauchy distributed, $\text{AsV}_{\hat{\lp}} (\omega)$ and $\text{AsV}_{\hat{\scpar}} (\omega)$ satisfy either condition (c1) or condition (c2). 

\begin{thm}
\label{thm:opt_omega}
If $\text{AsV}_{\hat{\lp}} (\omega)$ and $\text{AsV}_{\hat{\scpar}} (\omega)$ are differentiable  quasi-convex functions of $\omega$, then $\omega_{\gamma}^{*}$ lies in between the values of $\omega_{\lp}^{*}$ and $\omega_{\scpar}^{*}$.
\end{thm}

\begin{IEEEproof}
The proof is presented in Appendix \ref{app:proof_thm_opt_omega}. 
\end{IEEEproof}

In what follows, three sensing noise distributions, Gaussian, Laplace and Cauchy, are considered. In each case, (\ref{eqn:z_abs}) and (\ref{eqn:z_arg}) are applied to obtain the estimates, $\hat{\lp}$ and $\hat{\scpar}$, which are then used to construct $\hat{\gamma}$. The performance of all three schemes are studied and  $\omega_{\lp}^{*}$, $\omega_{\scpar}^{*}$ and $\omega_{\gamma}^{*}$ are also determined. 

\subsubsection{Gaussian Distribution}
\label{sssec:gaussian_tpc}

The case of Gaussian distributed sensing noise is considered first. The characteristic function in this case is given by
\begin{equation}
\varphi_{\eta} (\scpar \omega) = e^{-\omega^{2} \scpar^{2}/2}
\label{eqn:cf_gaussian}
\end{equation}
and the value of $\bar{z}$ is
\begin{equation}
\bar{z} = \sqrt{P} e^{j \omega \lp} e^{-\omega^{2} \scpar^{2}/2}. 
\label{eqn:z_gauss_pspc}
\end{equation}
The estimators using (\ref{eqn:z_abs}) and (\ref{eqn:z_arg}) are given by
\begin{align}
\hat{\lp} &= \frac{1}{\omega} \angle z_{L}, \label{eqn:lp_naive_gauss_pspc} \\
\hat{\scpar} &= \frac{1}{\omega} \sqrt{\log \left( \frac{P}{|z_{L}|^{2}} \right)}. \label{eqn:scp_naive_gauss_pspc}
\end{align}
The asymptotic variances are calculated to be
\begin{align}
\text{AsV}_{\hat{\lp}} (\omega) &= \frac{P+\sigma_{\nu}^{2}-P e^{-2 \omega^{2} \scpar^{2}}}{2P \omega^{2} e^{-\omega^{2} \scpar^{2}}} \label{eqn:asv_naive_tpc_gauss_lp} \\ 
\text{AsV}_{\hat{\scpar}} (\omega) &= \frac{P+\sigma_{\nu}^{2} - 2P e^{-\omega^{2} \scpar^{2}} + P e^{-2\omega^{2} \scpar^{2}}}{2P\omega^{4} \scpar^{2} e^{-\omega^{2} \scpar^{2}}}. 
\label{eqn:asv_naive_tpc_gauss_scpar}
\end{align}

The value of $\omega$ that minimizes the asymptotic variance of $\hat{\scpar}$ will now be computed. Making the substitution $\beta \leftarrow \omega^{2} \scpar^{2}$ and differentiating with respect to $\beta$, the following equation is required to be solved to find the stationary points of the asymptotic variance of $\hat{\scpar}$:
\begin{equation}
\beta\left[ e^{2\beta} \left(\frac{\sigma_{\nu}^{2}}{P} + 1\right) - 1 \right] - e^{2\beta} \left(\frac{\sigma_{\nu}^{2}}{P} + 1\right) + 2e^{\beta} - 1 = 0, 
\label{eqn:soln_min_asv_b}
\end{equation}
which depends on $\scpar$, $\sigma_{\nu}^{2}$ and $P$, but not on $\lp$. It is straightforward to show that in the Gaussian case, $\partial^{2} \text{AsV}_{\hat{\scpar}} (\omega)/\partial \omega^{2}>0$. Therefore, the asymptotic variance is convex, and the solution to (\ref{eqn:soln_min_asv_b}) leads to the unique minimum, $\omega_{\scpar}^{*} = \sqrt{\beta_{\scpar}^{\rm opt}}/\scpar$, where $\beta_{\scpar}^{\rm opt}$ is the solution to (\ref{eqn:soln_min_asv_b}).  Similarly, it can be shown that the asymptotic variance of $\hat{\lp}$ is convex. The value of $\omega$ that minimizes the asymptotic variance is given by $\omega_{\lp}^{*}= \sqrt{\beta_{\lp}^{\rm opt}}/\scpar$, where $\beta_{\lp}^{\rm opt}$ is the solution to
\begin{equation}
\left( \frac{\sigma_{\nu}^{2}}{P} + 1 \right) (\beta - 1) e^{2\beta} + (\beta + 1) = 0, 
\label{eqn:soln_min_asv_a}
\end{equation}
which depends on $\scpar$, $\sigma_{\nu}^{2}$ and $P$, but not on $\lp$. Neither (\ref{eqn:soln_min_asv_b}) nor (\ref{eqn:soln_min_asv_a}) can be solved analytically, but the solutions can be obtained numerically. 

The asymptotic variance of the SNR estimate is calculated using (\ref{eqn:SNR_asv}) and is given by
\begin{equation}
\text{AsV}_{\hat{\gamma}} (\omega) = 2 \gamma \frac{ \omega^{2} \left[ P + \sigma_{\nu}^{2} - 2 P e^{-2 \omega^{2} \scpar^{2}} \right] + \gamma \left[ P + \sigma_{\nu}^{2} - 2P e^{-\omega^{2} \scpar^{2}} + P e^{-2\omega^{2} \scpar^{2}} \right]}{ P \omega^{4} \scpar^{4} e^{-\omega^{2} \scpar^{2}}}. 
\label{eqn:gauss_asv_g}
\end{equation}
Since $\text{AsV}_{\hat{\lp}} (\omega)$ and $\text{AsV}_{\hat{\scpar}} (\omega)$ are convex, from Theorem \ref{thm:opt_omega}, $\omega_{\gamma}^{*}$ lies between the values of $\omega_{\lp}^{*}$ and $\omega_{\gamma}^{*}$. It is easy to verify that $\omega_{\gamma}^{*} = \sqrt{\beta_{\gamma}^{\rm opt}}/\scpar$, where $\beta_{\gamma}^{\rm opt}$ is the solution to
\begin{align}
\beta & \left[ \beta \left\{ e^{2\beta} \left( \frac{\sigma_{\nu}^{2}}{P} + 1 \right) + 1\right\} -e^{2\beta} \left( \frac{\sigma_{\nu}^{2}}{P} + 1 \right) + 1\right] \nonumber \\
+ &\gamma  \left[ \beta \left\{ e^{2\beta} \left( \frac{\sigma_{\nu}^{2}}{P} + 1 \right) -1 \right\} - 2 \left\{ e^{2\beta} \left( \frac{\sigma_{\nu}^{2}}{P} + 1 \right) - 2 e^{\beta} + 1 \right\} \right] =0. 
\label{eqn:soln_min_asv_g}
\end{align}

\subsubsection{Laplace Distribution}
\label{sssec:laplace_tpc}

Let $\eta_{i}$ be drawn from a Laplace distribution of mean zero and variance $\scpar^{2}$. The characteristic function is 
\begin{equation}
\varphi_{\eta}(\scpar \omega) = \frac{1}{1 + \frac{\omega^{2} \scpar^{2}}{2}}
\label{eqn:cf_laplace}
\end{equation}
and the value of $\bar{z}$ is
\begin{equation}
\bar{z} = \frac{\sqrt{P} e^{j \omega \lp}}{1 + \frac{\omega^{2} \scpar^{2}}{2}}. 
\label{eqn:z_laplace_pspc}
\end{equation}
The estimators in this case are
\begin{align}
\hat{\lp} &= \frac{1}{\omega} \angle z_{L},
\label{eqn:lp_naive_laplace_pspc}\\
\hat{\scpar} &= \frac{\sqrt{2}}{\omega} \sqrt{\frac{\sqrt{P}}{|z_{L}|} - 1},
\label{eqn:scp_naive_laplace_pspc}
\end{align}
with the asymptotic variances of $\hat{\lp}$ and $\hat{\scpar}$ given by
\begin{align}
\text{AsV}_{\hat{\lp}} (\omega) &= \frac{\left( \frac{\sigma_{\nu}^{2}}{P} + 1 \right) \left( 1 + 2 \omega^{2} \scpar^{2} \right) -1  }{8 \omega^{2} \left( 1+ 2 \omega^{2} \scpar^{2} \right) \left( 2 + \omega^{2} \scpar^{2} \right)^{-2}}, \label{eqn:asv_naive_tpc_laplace_lp} \\
\text{AsV}_{\hat{\scpar}} (\omega) &= \frac{\frac{\sigma_{\nu}^{2}}{P} \left( 2 + \scpar^{2} \omega^{2} \right)^{2} + \omega^{2} \scpar^{2} \left( 6 + \omega^{2} \scpar^{2} \right)} {32 \omega^{4} \scpar^{2} \left( 2 + \omega^{2} \scpar^{2} \right)^{-2}}. 
\label{eqn:ascov_naive_tpc_laplace_scpar}
\end{align}
Using (\ref{eqn:SNR_asv}), the asymptotic variance of $\hat{\gamma}$ is given by
\begin{align}
\text{AsV}_{\hat{\gamma}} (\omega) =& \frac{\gamma \left( 2 + \omega^{2} \scpar^{2} \right)^{2}}{8 P \omega^{4} \scpar^{4} \left( 1 + 2 \omega^{2} \scpar^{2} \right)}  \Big[ 4 \scpar^{2} \omega^{2} \left\{ 2 P \omega^{2} \scpar^{2} + \sigma_{\nu}^{2} \left( 1 + 2 \omega^{2} \scpar^{2} \right) \right\} \nonumber \\
& \phantom{abdef} + \gamma \left( 1 + 2 \omega^{2} \scpar^{2} \right) \left\{ P \scpar^{2} \omega^{2} \left( 6 + \omega^{2} \scpar^{2} \right) + \sigma_{\nu}^{2} \left( 2 + \omega^{2} \scpar^{2} \right)^2 \right\} \Big]. 
\label{eqn:asv_gamma_laplace}
\end{align}

To minimize the asymptotic variance of $\hat{\lp}$ it can be shown that $\omega_{\lp}^{*}$ is given by $\omega_{\lp}^{*}= \sqrt{\beta_{\lp}^{\rm opt}}/\scpar$, where \cite{robust_est_cihan_adarsh_2010}
\begin{equation}
\beta_{\lp}^{\rm opt} = \frac{1}{12} \left( \frac{c}{\frac{\sigma_{\nu}^{2}}{P} + 1} + \frac{25 \frac{\sigma_{\nu}^{2}}{P} + 4}{c} + 2 \right),
\label{eqn:soln_min_asv_a_laplace}
\end{equation}
and
\begin{equation}
c^{3} =  125 \left( \frac{\sigma_{\nu}^{2}}{P} \right)^{3} + 258 \left( \frac{\sigma_{\nu}^{2}}{P} \right)^{2} + 141 \left( \frac{\sigma_{\nu}^{2}}{P} \right) + 3 \sqrt{3} \sqrt{\left( \frac{\sigma_{\nu}^{2}}{P} \right) \left( \frac{\sigma_{\nu}^{2}}{P} + 1 \right)^{3} \left( 375 \frac{\sigma_{\nu}^{2}}{P} + 32\right)} + 8.
\label{eqn:c_laplace}
\end{equation}
To minimize the asymptotic variance of $\hat{\scpar}$, one needs to calculate $\omega_{\scpar}^{*} = \sqrt{\beta_{\scpar}^{\rm opt}}/\scpar$, where $\beta_{\scpar}^{\rm opt}$ is the solution to the quintic equation
\begin{equation}
16 \left( \frac{\sigma_{\nu}^{2}}{P} + 1 \right) \beta^{5} + 2 \left( 12 \frac{\sigma_{\nu}^{2}}{P} + 13 \right) \beta^{4} - \left( 7 \frac{\sigma_{\nu}^{2}}{P} + 8 \right) \beta^{3} -23 \frac{\sigma_{\nu}^{2}}{P} \beta^{2} - 9 \frac{\sigma_{\nu}^{2}}{P} \beta - \frac{\sigma_{\nu}^{2}}{P} = 0. 
\label{eqn:soln_min_asv_b_laplace}
\end{equation}
Similarly, the asymptotic variance of $\hat{\gamma}$ is minimized at $\omega_{\gamma}^{*} = \sqrt{\beta_{\gamma}^{\rm opt}}/\scpar$, where $\beta_{\gamma}^{\rm opt}$ is the solution to 
\begin{align}
16 & \left(\gamma + 2 + \gamma \frac{\sigma_{\nu}^{2}}{P} + 2 \frac{\sigma_{\nu}^{2}}{P} \right) \beta^{5} + 2 \left( 13 \gamma - 8 + 12 \gamma \frac{\sigma_{\nu}^{2}}{P} - 8 \frac{\sigma_{\nu}^{2}}{P} \right) \beta^{4} \nonumber \\
& + 7 \left( 7 \gamma \frac{\sigma_{\nu}^{2}}{P} - 14 \frac{\sigma_{\nu}^{2}}{P} - 8 \gamma \right) \beta^{3} - \left(23 \gamma + 2 \right) \frac{\sigma_{\nu}^{2}}{P} \beta^{2} -9 \gamma \frac{\sigma_{\nu}^{2}}{P} \beta - \gamma \frac{\sigma_{\nu}^{2}}{P} = 0. 
\label{eqn:soln_min_asv_g_laplace}
\end{align}
The quintic equations in (\ref{eqn:soln_min_asv_b_laplace}) and (\ref{eqn:soln_min_asv_g_laplace}) cannot be solved analytically. However, the solutions to these can be obtained numerically. It is straightforward to see that $\partial^{2} \text{AsV}_{\hat{\lp}} (\omega)/\partial \omega^{2} \geq 0$ and $\partial^{2} \text{AsV}_{\hat{\scpar}} (\omega)/\partial \omega^{2} \geq 0$, and therefore, $\text{AsV}_{\hat{\lp}} (\omega)$ and $\text{AsV}_{\hat{\scpar}} (\omega)$ are quasi-convex \cite[pp. 101]{boyd2004}. Therefore, from Theorem \ref{thm:opt_omega}, $\omega_{\gamma}^{*}$ lies between $\omega_{\lp}^{*}$ and $\omega_{\gamma}^{*}$. 

\subsubsection{Cauchy Distribution}
\label{sssec:cauchy_tpc}

Since the Cauchy distribution does not have any finite moments, the scale parameter in this case is selected to be the Cauchy parameter, $\scpar$, which specifies the half-width at half-maximum (HWHM) \cite{Gillespie1992}. The characteristic function is given by
\begin{equation}
\varphi_{\eta}(\scpar \omega) = e^{-\scpar\omega}, \quad \omega > 0, 
\label{eqn:cauchy_char_fn}
\end{equation}
to yield
\begin{equation}
\bar{z} = \sqrt{P} e^{j \omega \lp} e^{-\omega \scpar}
\label{eqn:z_cauchy_pspc}
\end{equation} 
and the estimates of $\lp$ and $\scpar$ are given as
\begin{align}
\hat{\lp} &= \frac{1}{\omega} \angle z_{L},
\label{eqn:lp_naive_cauchy_pspc}\\
\hat{\scpar} &= \frac{1}{\omega} \log \left( \frac{\sqrt{P}}{|z_{L}|} \right).
\label{eqn:scp_naive_cauchy_pspc}
\end{align}
These estimators have the asymptotic variance given by
\begin{equation}
\text{AsV}_{\hat{\lp}} (\omega) = \text{AsV}_{\hat{\scpar}} (\omega) = \frac{P + \sigma_{\nu}^{2} - P e^{-2 \omega \scpar}} {2 P \omega^{2} e^{-2\omega\scpar}}.
\label{eqn:ascov_naive_cauchy_tpc}
\end{equation}
The asymptotic variance of $\hat{\gamma}$ can be calculated using (\ref{eqn:SNR_asv}) and is given by
\begin{equation}
\text{AsV}_{\hat{\gamma}} (\omega) = \frac{2 \gamma \left( \gamma + 1 \right)\left( P + \sigma_{\nu}^{2} - P e^{-2 \omega \scpar}\right)}{P \omega^{2} \scpar^{2} e^{-2\omega\scpar}}. 
\label{eqn:asv_gamma_cauchy}
\end{equation}

Since the asymptotic variances of both $\hat{\lp}$ and $\hat{\scpar}$ are identical, and can be shown to be quasi-convex, from Theorem \ref{thm:opt_omega}, the same value of $\omega$ minimizes the asymptotic variances of all $\hat{\lp}$, $\hat{\scpar}$ and $\hat{\gamma}$. Taking the first derivative of the asymptotic variance with respect to $\omega$ and equating to zero, the value of $\omega$ that minimizes the asymptotic variances is given by
\begin{equation}
\omega^{*} = \frac{2 + W \left[ - \frac{2P}{e^{2} \left( P + \sigma_{\nu}^{2} \right)} \right]}{2 \scpar},
\label{eqn:soln_min_asv_cauchy}
\end{equation}
where $W(\cdot)$ is the Lambert-$W$ function, which is the inverse function of $x e^{x}$ \cite{Lambert1996}. 

\section{Per-Sensor Power Constraint}
\label{sec:pspc}

In the case of per-sensor power constraint, the total transmit power increases as the number of sensors in the system increases, with the channel noise variance remaining the same. Each sensor transmits with a power of $P$ and the signal at the FC, shown in (\ref{eqn:yn}) is given by
\begin{equation}
y_{L} = \sqrt{P} \sum_{l=1}^{L} e^{j \omega x_{l}} + \nu. 
\label{eqn:yn_pspc}
\end{equation}

\subsection{The Estimator}
\label{ssec:estimator_pspc}

At the FC, the signal from (\ref{eqn:yn_pspc}) is modified to give
\begin{equation}
\zeta_{L} \mathop{:=} \frac{y_{L}}{L} = \sqrt{P} \frac{1}{L} \sum_{i=1}^{L} e^{j \omega x_{i}} + \frac{\nu}{L},
\label{eqn:zeta_L}
\end{equation}
which as $L\to\infty$, converges with probability one to
\begin{equation}
\zeta = \lim_{L\to\infty} \zeta_{L} =  \sqrt{P} \lim_{L\to\infty} \frac{1}{L} \sum_{i=1}^{L} e^{j \omega x_{i}} = \sqrt{P} e^{j \omega \lp} \varphi_{\eta} (\scpar \omega). 
\label{eqn:zeta}
\end{equation}
Defining $\boldsymbol{\zeta}_{L} = [\zeta_{L}^{R} \text{  } \zeta_{L}^{I}]$ and $\boldsymbol{\zeta} = [\zeta^{R} \text{  } \zeta^{I}]$, $\boldsymbol{\zeta}_{L}$ converges to $\boldsymbol{\zeta}$ in such a way that
\begin{equation}
\tilde{\boldsymbol{\zeta}} = \lim_{L\to\infty} \sqrt{L} (\boldsymbol{\zeta}_{L} - \boldsymbol{\zeta})
\label{eqn:zeta_conv}
\end{equation}
is a $2\times 1$ Gaussian random vector with zero mean and a $2\times 2$ covariance matrix $\boldsymbol{\tilde{\boldsymbol{\Sigma}}}$ with elements
\begin{align}
\tilde{\Sigma}_{11} &= P \left[ \tilde{v}_{c} \cos^{2}(\omega \lp) + \tilde{v}_{s} \sin^{2}(\omega \lp) \right] \nonumber\\
\tilde{\Sigma}_{22} &= P \left[ \tilde{v}_{s} \cos^{2}(\omega \lp) + \tilde{v}_{c} \sin^{2}(\omega \lp) \right] \nonumber\\
\tilde{\Sigma}_{12} &= \tilde{\Sigma}_{21} = P (\tilde{v}_{c} - \tilde{v}_{s})
\sin(\omega \lp) \cos(\omega \lp),
\label{eqn:sigma_defn_pspc}
\end{align}
where the parameters $\tilde{v}_{c} \mathop{:=} \text{var} [\cos (\omega \eta_{l})] = 1/2 + \varphi_{\eta}(2 \scpar \omega) /2 - \varphi_{\eta}^{2}(\scpar \omega)$ and $\tilde{v}_{s} \mathop{:=} \text{var}[\sin (\omega \eta_{l})] = 1/2 - \varphi_{\eta}(2 \scpar \omega)/2$. The minimum variance estimator for $[\hat{\lp} \text{  } \hat{\scpar}]^{T}$ in this case is given by
\begin{equation}
\left[ \hat{\lp} \quad \hat{\scpar} \right]^{T} = \argmin_{\lp, \scpar} [\boldsymbol{\zeta}_{L} -
\boldsymbol{\zeta}]^{T} \boldsymbol{\tilde{\boldsymbol{\Sigma}}}^{-1} 
[\boldsymbol{\zeta}_{L} - \boldsymbol{\zeta}],
\label{eqn:est_lp_scpar_pspc}
\end{equation}
and the asymptotic covariance matrix of the estimates is given by $\left[ \mathbf{J}^{T} \boldsymbol{\tilde{\boldsymbol{\Sigma}}}^{-1} \mathbf{J} \right]^{-1}$ to yield the asymptotic variances
\begin{align}
\text{AsV}_{\hat{\lp}} (\omega) &= \frac{1-\varphi_{\eta} (2 \scpar \omega)}{2 \omega^{2} \varphi_{\eta}^{2} (\scpar \omega)} \label{eqn:asv_lp_pspc} \\ \text{AsV}_{\hat{\scpar}} (\omega) &= \frac{1-2\varphi_{\eta}^{2} (\scpar \omega) + \varphi_{\eta} (2 \scpar \omega)}{2 \left[ \frac{\partial \varphi_{\eta}(\scpar \omega)}{\partial \scpar} \right]^{2}}, \label{eqn:asv_scpar_pspc}
\end{align}
which can be verified to be (\ref{eqn:asv_lp_tpc}) and (\ref{eqn:asv_scpar_tpc}) with $\sigma_{\nu} \to 0$. 

The development in this per-sensor power constraint case shows that as the number of sensors increases, the effect of channel noise becomes negligible. In fact, the results in the case of per-sensor power constraint can be interpreted as a special case of the results in Section \ref{sec:tpc}, with $\sigma_{\nu}^{2}\to 0$. These results are separately presented since closed form solutions can be obtained for $\omega^{*}$ for the different sensing noise distributions considered. The estimate of the SNR is computed as given in (\ref{eqn:SNR_est_defn}), with asymptotic variance as given in (\ref{eqn:SNR_asv}). Theorem \ref{thm:compare_estimators}, Theorem \ref{thm:opt_omega} and the ML invariance property continue to hold. 

The three sensing noise distributions considered previously, the Gaussian distribution, the Laplace distribution and the Cauchy distribution are considered again for the per-sensor power constraint case. In each case, the performance is evaluated and the values of $\omega$ that minimize the asymptotic variances of $\hat{\lp}$, $\hat{\scpar}$ and $\hat{\gamma}$ are calculated. 

\subsubsection{Gaussian Distribution}
\label{sssec:gauss_pspc}

The performance in this case is given by substituting $\sigma_{\nu} = 0$ in (\ref{eqn:asv_lp_tpc}) and (\ref{eqn:asv_scpar_tpc}) to give
\begin{align}
\text{AsV}_{\hat{\lp}} (\omega) &= \frac{1-e^{-2\omega^{2} \scpar^{2}}}{2\omega^{2}e^{-\omega^{2} \scpar^{2}}} \label{eqn:asv_naive_pspc_gauss_lp}, \\ \text{AsV}_{\hat{\scpar}} (\omega) &= \frac{\left( 1 - e^{-\omega^{2} \scpar^{2}} \right)^{2}}{2\omega^{4} \scpar^{2} e^{-\omega^{2} \scpar^{2}}}. \label{eqn:asv_naive_pspc_gauss_scpar}
\end{align}
The asymptotic variance of $\hat{\gamma}$ is given by
\begin{equation}
\text{AsV}_{\hat{\gamma}} (\omega) = \frac{\gamma \left( 1 - 2 e^{-\omega^{2} \scpar^{2}} + e^{-2 \omega^{2} \scpar^{2}} \right) + \omega^{2} \scpar^{2} \left( 1 - e^{-2 \omega^{2} \scpar^{2}} \right)}{2 \omega^{4} \scpar^{2} e^{-\omega^{2} \scpar^{2}}}.
\label{eqn:asv_gamma_gauss_pspc}
\end{equation}
The value of $\omega$ that minimizes the asymptotic variance of $\hat{\lp}$ is given by
\begin{equation}
\omega_{\lp}^{*} = \argmin_{\omega} \frac{1-e^{-2\omega^{2} \scpar^{2}}}{2\omega^{2}e^{-\omega^{2} \scpar^{2}}}. 
\label{eqn:opt_omega_lp_gauss}
\end{equation}
It can easily be verified that the objective is minimized as $\omega_{\lp}^{*}\to0$. In a similar way, it can be shown that $\omega_{\scpar}^{*}\to0$ minimizes the asymptotic variance of $\hat{\scpar}$, and that $\text{AsV}_{\hat{\gamma}} (\omega)$ is also minimized when $\omega_{\gamma}^{*}\to0$. However, if $\omega = 0$, the transmissions from the sensors do not depend on the sensed data, invalidating the choice of $\omega = 0$. Therefore, in order to minimize the asymptotic variance, a sufficiently small value of $\omega$ is selected at the sensors \cite{robust_est_cihan_adarsh_2010}. The apparent discrepancy between the choice of $\omega$ suggested by the asymptotic analysis (small $\omega$) and its limiting value of $\omega = 0$ is due to the asymptotic nature of our analysis. In Figure \ref{Fig:gauss_pspc_diff_L}, we show that as $\omega \to 0$ for small $L$, the value of $\text{AsV}_{\hat{\lp}} (\omega)$ increases greatly, indicating poor performance, as expected. However, for large $L$, the minimum is for smaller values of $\omega$. We elaborate more on Figure \ref{Fig:gauss_pspc_diff_L} in Section \ref{sec:simulations}. 

\subsubsection{Laplace Distribution}
\label{sssec:laplace_pspc}

The asymptotic variances are given by (\ref{eqn:asv_lp_tpc}) and (\ref{eqn:asv_scpar_tpc}) with $\sigma_{\nu} \to 0$:
\begin{align}
\text{AsV}_{\hat{\lp}} (\omega) &= \frac{\scpar^{2} \left( 2 + \omega^{2} \scpar^{2} \right)^{2}} {4 \left( 1 + 2 \omega^{2} \scpar^{2} \right)} \label{eqn:asv_naive_pspc_laplace_lp}, \\
\text{AsV}_{\hat{\scpar}} (\omega) &= \frac{ \scpar^{2} \left( 2 + \omega^{2} \scpar^{2} \right)^{2} \left( 5 + \omega^{2} \scpar^{2} \right)} {16 \left( 1 + 2 \omega^{2} \scpar^{2} \right)}. \label{eqn:asv_naive_pspc_laplace_scpar}
\end{align}
The asymptotic variance of $\hat{\gamma}$ is given by
\begin{equation}
\text{AsV}_{\hat{\gamma}} (\omega) = \frac{\gamma \left( 2 + \omega^{2} \scpar^{2} \right)^{2} \left[ 8 \omega^{2} \scpar^{2} + \gamma \left( 1 + 2 \omega^{2} \scpar^{2} \right) \left( 6 + \omega^{2} \scpar^{2} \right) \right] }{8 \omega^{2} \scpar^{2} \left( 1 + 2 \omega^{2} \scpar^{2} \right)}. 
\label{eqn:asv_gamma_laplace_pspc}
\end{equation}

To identify the value of $\omega$ that yields the best performance for estimating $\lp$, the following problem needs to be solved:
\begin{equation}
\omega_{\lp}^{*} = \argmin_{\omega}  \frac{\scpar^{2} \left( 2 + \omega^{2} \scpar^{2} \right)^{2}} {4 \left( 1 + 2 \omega^{2} \scpar^{2} \right)}.
\label{eqn:opt_omega_lp_laplace}
\end{equation}
By inspecting the first derivative, it can be verified that $\omega_{\lp}^{*} = 1/\scpar$. For the case of $\hat{\scpar}$
\begin{equation}
\omega_{\scpar}^{*} = \argmin_{\omega} \frac{\left( 2 + \omega^{2} \scpar^{2} \right)^{2} \left( 6 + \omega^{2} \scpar^{2} \right)}{32 \omega^{2}}. 
\label{eqn:opt_omega_scp_laplace}
\end{equation}
This is minimized at $\omega_{\scpar}^{*} = (1/\scpar \sqrt{8})\sqrt{3\sqrt{33} - 13} > \omega_{\lp}^{*}$. The value of $\omega$ that minimizes $\text{AsV}_{\hat{\gamma}} (\omega)$ is similarly calculated to be
\begin{equation}
\omega_{\gamma}^{*} = \frac{\sqrt{-13 \gamma - 16 + \sqrt{\left( 9 \gamma + 16 \right) \left( 33 \gamma + 16 \right)}}}{4 \scpar \sqrt{\gamma}}. 
\label{eqn:opt_omega_gamma_laplace}
\end{equation}

\subsubsection{Cauchy Distribution}
\label{sssec:cauchy_pspc}

In the case of Cauchy distributed sensing noise, the asymptotic variances for the estimates, $\hat{\lp}$, $\hat{\scpar}$ and $\hat{\gamma}$, are given by
\begin{align}
\text{AsV}_{\hat{\lp}} (\omega) &= \text{AsV}_{\hat{\scpar}} (\omega) = \frac{1-e^{-2\omega\scpar}}{2\omega^{2} e^{-2 \omega \scpar}},
\label{eqn:asv_naive_cauchy_pspc} \\
\text{AsV}_{\hat{\gamma}} (\omega) &= \frac{2\gamma \left( \gamma + 1 \right) \left( 1 - e^{2 \omega \scpar} \right)}{\omega^{2} \scpar^{2} e^{-2 \omega \scpar}}.
\label{eqn:asv_gamma_cauchy_pspc}
\end{align}
Since $\text{AsV}_{\hat{\lp}} (\omega)$ and $\text{AsV}_{\hat{\scpar}} (\omega)$ are identical, the value of $\omega$ that minimizes them is the same. Therefore, the value of $\omega$ that minimizes $\text{AsV}_{\hat{\lp}} (\omega)$, $\text{AsV}_{\hat{\scpar}} (\omega)$ and $\text{AsV}_{\hat{\gamma}} (\omega)$ is given by 
\begin{equation}
\omega^{*} = \frac{2+W(-2e^{-2})} {2\scpar}.
\label{eqn:opt_omega_lp_cauchy}
\end{equation}

\subsection{Asymptotic Relative Efficiency}
\label{sec:are}

Since a non-linear scheme is used to transmit the observations from the sensors to the FC, it is required to evaluate the loss of information due to this processing. To evaluate this loss, the asymptotic relative efficiency is defined for each of the estimators. 

In the case of estimating $\scpar$, the asymptotic relative efficiency is defined as 
\begin{equation}
\mathcal{E}_{\hat{\scpar}} = \left[ I_{\scpar} \inf_{\omega\in (0,2\pi/\lp_{R}]} \text{AsV}_{\hat{\scpar}} (\omega) \right]^{-1},
\label{eqn:are_scpar}
\end{equation}
where $I_{\scpar}$ is the Fisher information of the observations, $x_{i}$, about the parameter $\scpar$ \cite{Huber1981, Lehmann1998, Johnson2004}, and $\text{AsV}_{\hat{\scpar}} (\omega)$ is the asymptotic variance of the estimator of $\scpar$. 
It can be shown that $I_{\scpar}$ depends only on $\scpar$ and not on $\lp$, but $\mathcal{E}_{\hat{\scpar}}$ is independent of both $\scpar$ and $\lp$. The asymptotic relative efficiency depends only on the distribution of the sensing noise and $0 \leq \mathcal{E}_{\hat{\scpar}} \leq 1$. 

\begin{table}[tb]
\begin{centering}
\begin{tabular}{|c||c|c|c|}
\hline Distribution & Gaussian & Laplace & Cauchy\\
\hline $\mathcal{E}_{\hat{\lp}}$ & $1$ & $0.66$ & $0.65$\\
\hline $\mathcal{E}_{\hat{\scpar}}$ & $1$ & $0.5$ & $0.65$ \\
\hline
\end{tabular}
 \caption{Information Efficiency for $\lp$ and $\scpar$ different distributions. }
 \label{table:info_eff_values}
\end{centering}
\end{table}

The second row of Table \ref{table:info_eff_values} shows the values of the asymptotic relative efficiency when estimating $\scpar$ for the Gaussian, Laplace and Cauchy distributions. The Fisher information is calculated using the definitions in \cite{Huber1981, Lehmann1998, Johnson2004}, for the different distributions, yielding $2 \scpar^{-2}$, $\scpar^{-2}$ and $0.5\scpar^{-2}$ for the Gaussian, Laplace and Cauchy distributions, respectively. The values of $\inf_{\omega\in (0,2\pi/\lp_{R}]} \text{AsV}_{\hat{\scpar}} (\omega)$ are calculated using (\ref{eqn:asv_naive_pspc_gauss_scpar}) for the Gaussian distribution, (\ref{eqn:asv_naive_pspc_laplace_scpar}) for the Laplace distribution, and (\ref{eqn:asv_naive_cauchy_pspc}) for the Cauchy distribution. When estimating $\scpar$, for the distributions considered in this paper (Gaussian, Laplace and Cauchy), the asymptotic relative efficiency is one only in the case when the sensing noise is Gaussian. 

Similarly, in the case of estimating $\lp$, the asymptotic relative efficiency is defined as 
\begin{equation}
\mathcal{E}_{\hat{\lp}} = \left[ I_{\lp} \inf_{\omega\in (0,2\pi/\lp_{R}]} \text{AsV}_{\hat{\lp}} (\omega) \right]^{-1},
\label{eqn:are_lp}
\end{equation}
where $I_{\lp}$ is the Fisher information of the observations, $x_{i}$, about the parameter $\lp$ \cite{Huber1981, Lehmann1998, Johnson2004}, and $\text{AsV}_{\hat{\lp}} (\omega)$ is the asymptotic variance of the estimator of $\lp$. It is well known that $I_{\lp}$ is independent of $\lp$. The asymptotic relative efficiency for the estimators of $\lp$ are recorded in the first row of Table \ref{table:info_eff_values}. The values of the Fisher information are given by $\scpar^{-2}$, $2 \scpar^{-2}$ and $0.5 \scpar^{-2}$ for the Gaussian, Laplace and Cauchy distributions, respectively, and the values of $\inf_{\omega\in (0,2\pi/\lp_{R}]} \text{AsV}_{\hat{\lp}} (\omega)$ are calculated using (\ref{eqn:asv_naive_pspc_gauss_lp}), (\ref{eqn:asv_naive_pspc_laplace_lp}) and (\ref{eqn:asv_naive_cauchy_pspc}), respectively. 

It can be seen from these results that there is no loss in efficiency only in the case of the Gaussian distribution. It has been shown in \cite{tepedelenlioglu2010} that in the case of estimating $\lp$, the asymptotic relative efficiency is one if and only if the noise distribution is Gaussian. In all other cases, information is lost due to the non-linear processing at the sensors. 

Note that these results do not indicate that Gaussian noise yields the best performance. It means that only in the case of Gaussian sensing noise, there is asymptotically no loss in information about $\lp$ or $\scpar$ due to the transformation $x\rightarrow e^{j \omega x}$. As an example, consider estimating $\lp$, where $\{ \eta_{i} \}$ are Laplace distributed. It can be shown that $\inf_{\omega\in (0,2\pi/\lp_{R}]} \text{AsV}_{\hat{\lp}} (\omega)$ is smaller for the Laplace sensing noise compared to the Gaussian case \cite{tepedelenlioglu2010}. 

\section{Simulation Results}
\label{sec:simulations}

\begin{figure}[tb]
\begin{minipage}[b]{1.0\linewidth}
  \centering
  \centerline{\epsfig{figure=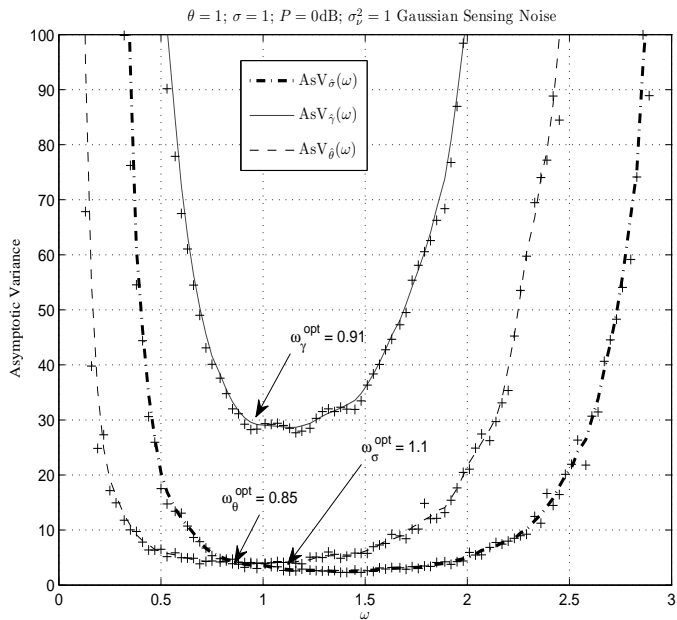,width=10.5cm,height=9cm}}
\end{minipage}
\caption{Asymptotic variance vs. scale parameter. Sensing noise is Gaussian distributed. The asymptotic variances match the CRLB.} \label{Fig:gauss_tpc}
\end{figure}

\begin{figure}[tb]
\begin{minipage}[b]{1.0\linewidth}
  \centering
  \centerline{\epsfig{figure=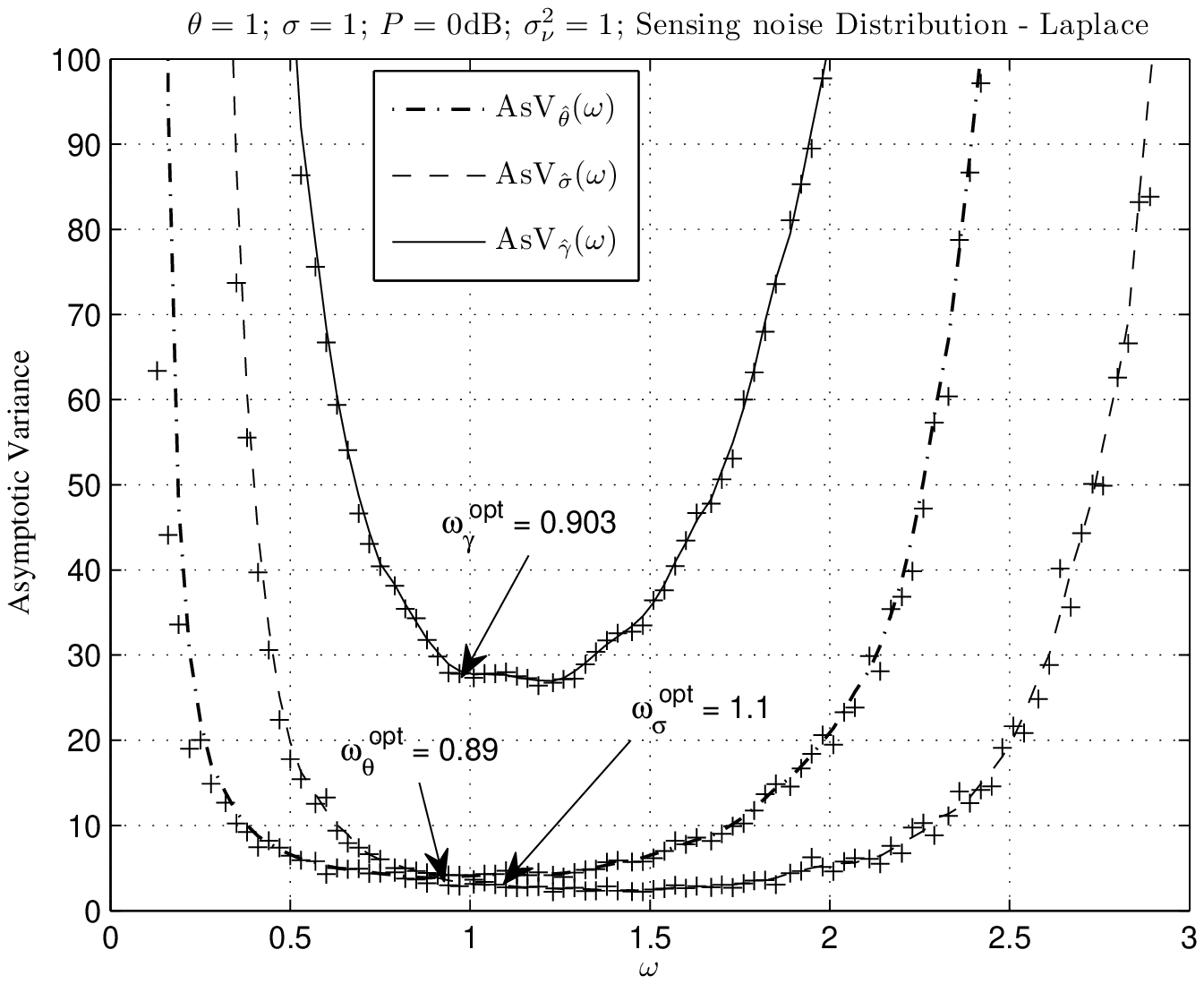,width=10.5cm,height=9cm}}
\end{minipage}
\caption{Asymptotic variance vs. scale parameter. Sensing noise is Laplace distributed. The asymptotic variances match the CRLB.} \label{Fig:laplace_tpc}
\end{figure}

\begin{figure}[tb]
\begin{minipage}[b]{1.0\linewidth}
  \centering
  \centerline{\epsfig{figure=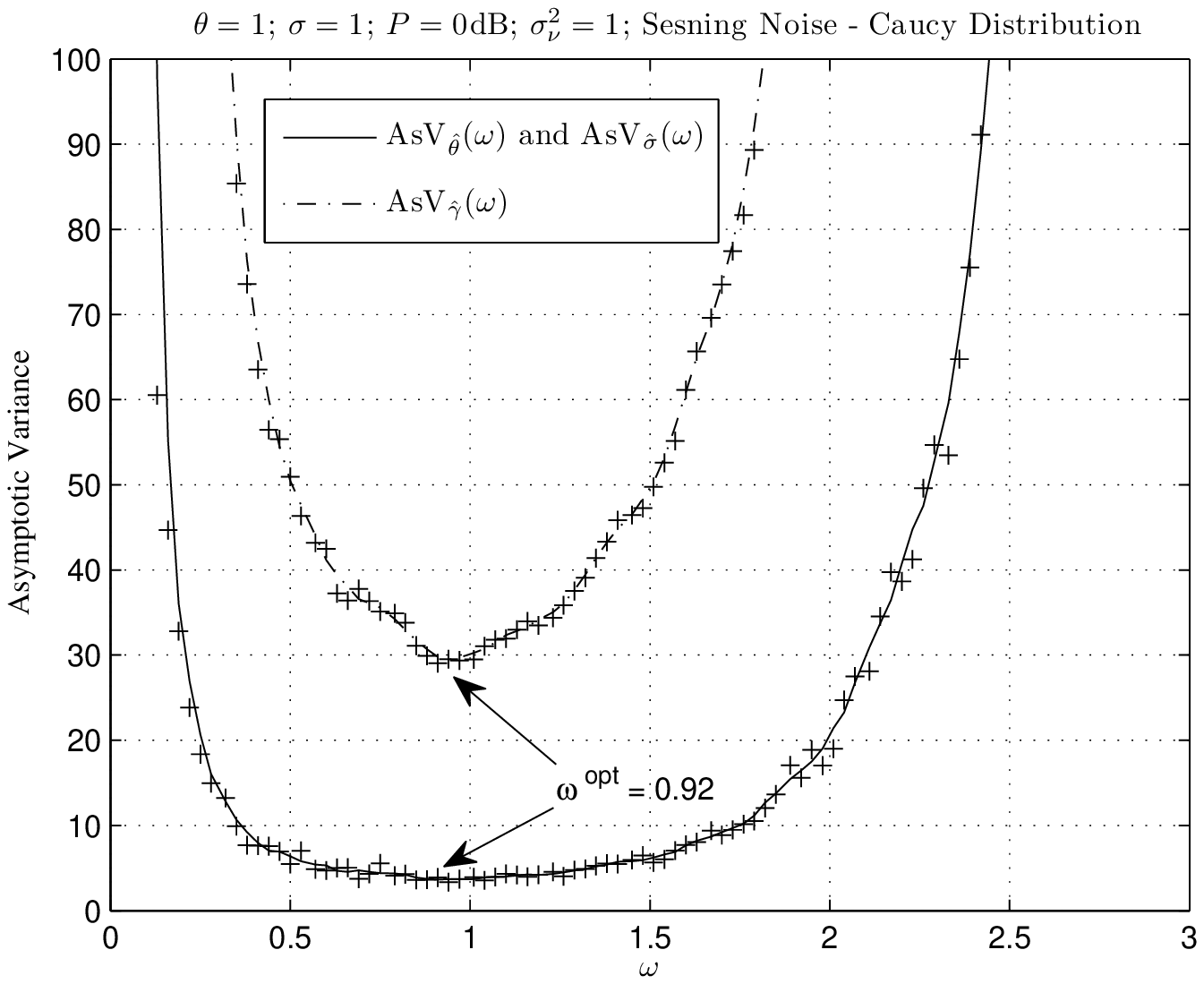,width=10.5cm,height=9cm}}
\end{minipage}
\caption{Asymptotic variance vs. scale parameter. Sensing noise is Cauchy distributed. The asymptotic variances match the CRLB.} \label{Fig:cauchy_tpc}
\end{figure}

\begin{figure}[tb]
\begin{minipage}[b]{1.0\linewidth}
  \centering
  \centerline{\epsfig{figure=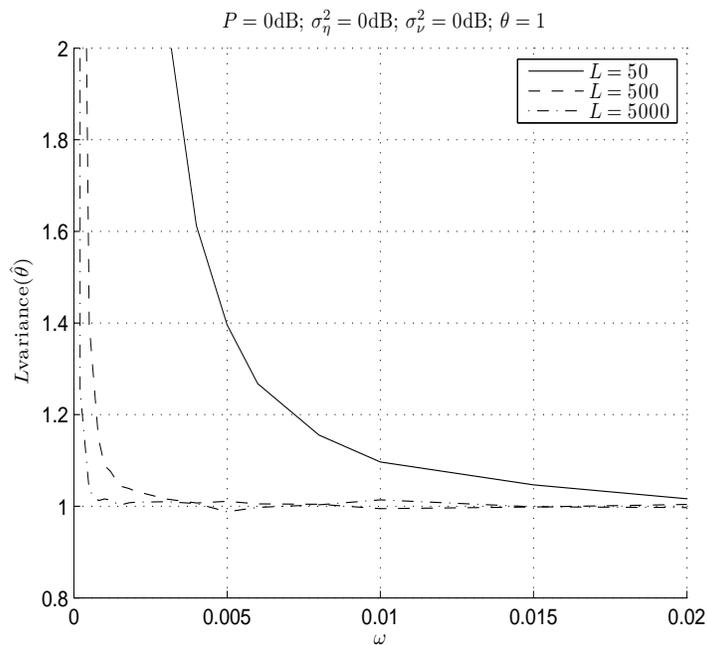,width=10.5cm,height=9cm}}
\end{minipage}
\caption{Plot of $L$variance$(\hat{\lp})$ vs. $\omega$ for different values of $L$.} \label{Fig:gauss_pspc_diff_L}
\end{figure}

\begin{figure}[tb]
\begin{minipage}[b]{1.0\linewidth}
  \centering
  \centerline{\epsfig{figure=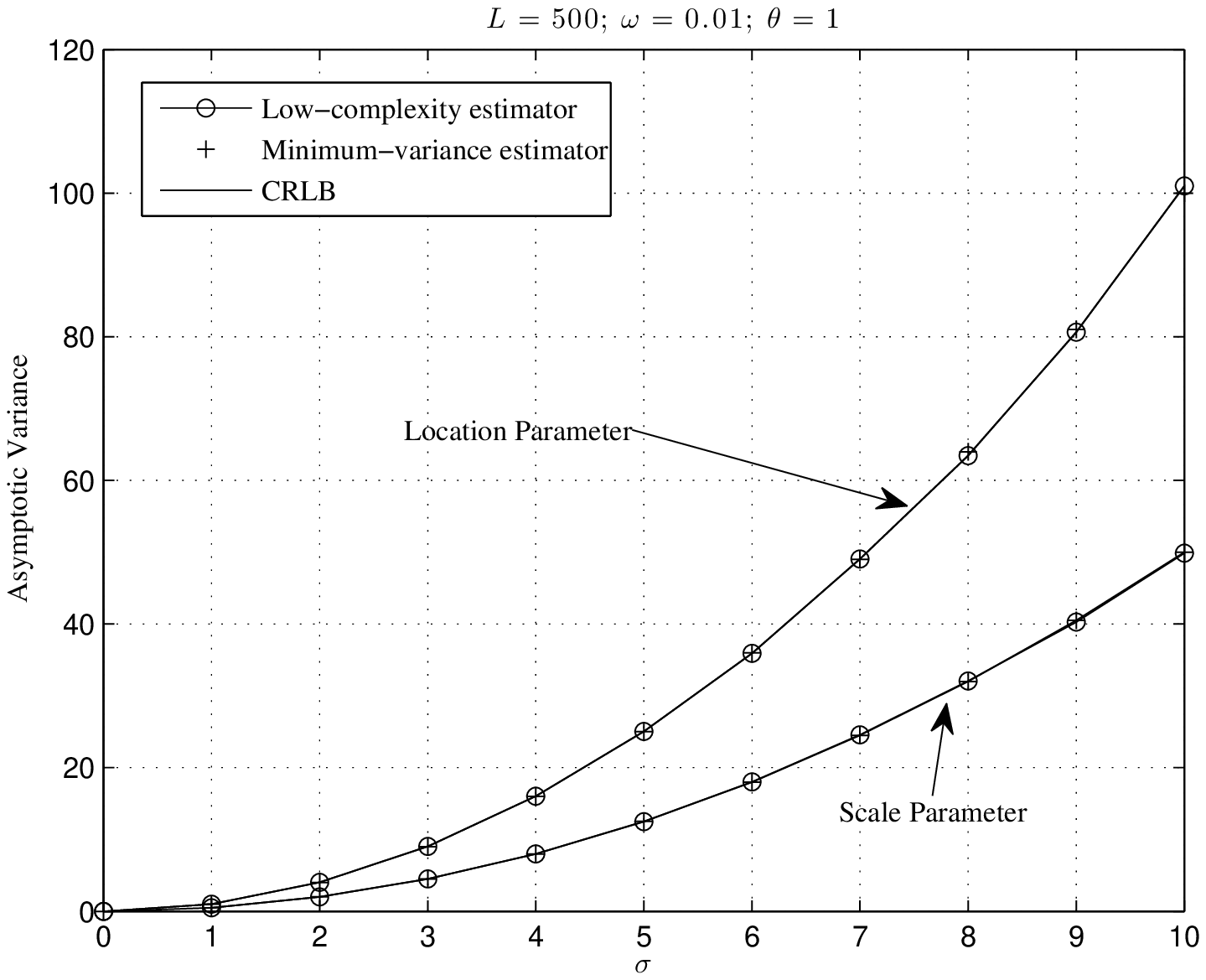,width=10.5cm,height=9cm}}
\end{minipage}
\caption{Asymptotic variance vs. $\omega$. Sensing noise is Gaussian distributed. Simulation values match numerical results} \label{Fig:gauss}
\end{figure}

Simulations are first carried out for the total power constraint case, and then for the per-sensor power constraint case. For the total power constraint case, in each case of sensing distribution, the estimators are simulated and compared. The values of the optimum $\omega$ for each parameter are indicated on the graphs and compared with the theoretical values. In the case of per-sensor power constraint, the estimators are evaluated across different values of  $\scpar$. The performance measures are compared against the inverse of the Fisher Information, which is known to be the Cram\'{e}r-Rao Lower Bound (CRLB), a lower bound on the variance of unbiased estimators. 

Figure \ref{Fig:gauss_tpc} shows the performance of the estimators of $\lp$, $\scpar$ and $\gamma$ versus $\omega$ in the total power constraint case and Gaussian sensing noise. It can be seen from the plots that $\omega_{\lp}^{*} \leq \omega_{\gamma}^{*} \leq \omega_{\scpar}^{*}$ as expected. The values of $\omega_{\lp}^{*}$, $\omega_{\scpar}^{*}$ and $\omega_{\gamma}^{*}$ are also calculated from (\ref{eqn:soln_min_asv_a}), (\ref{eqn:soln_min_asv_b}) and (\ref{eqn:soln_min_asv_g}), respectively, and marked on the figure, verifying the results. 

The system with Laplace sensing noise is simulated and the results are shown in Figure \ref{Fig:laplace_tpc}. The estimators of $\lp$, $\scpar$ and $\gamma$ are evaluated and the performance is plotted versus $\omega$ when the total power is constrained across the sensors. As expected from the results in Section \ref{sssec:laplace_tpc}, $\omega_{\lp}^{*} \leq \omega_{\gamma}^{*} \leq \omega_{\scpar}^{*}$. Using the formulas in (\ref{eqn:soln_min_asv_a_laplace}), (\ref{eqn:soln_min_asv_b_laplace}) and (\ref{eqn:soln_min_asv_g_laplace}), the values of $\omega_{\lp}^{*}$, $\omega_{\scpar}^{*}$ and $\omega_{\gamma}^{*}$ are calculated and shown on Figure \ref{Fig:laplace_tpc}, where the theoretical and simulation values are seen to agree. 

Figure \ref{Fig:cauchy_tpc} shows the performance of the system when the sensing noise is Cauchy distributed for the total power constraint case. The values of ${\rm AsV}_{\hat{\lp}} (\omega)$, ${\rm AsV}_{\hat{\scpar}} (\omega)$ and ${\rm AsV}_{\hat{\gamma}} (\omega)$ are plotted against $\omega$. It is easily seen that ${\rm AsV}_{\hat{\lp}} (\omega) = {\rm AsV}_{\hat{\scpar}} (\omega)$ and it is verified that $\omega_{\lp}^{*} = \omega_{\scpar}^{*} = \omega_{\gamma}^{*}$. The theoretical value from (\ref{eqn:soln_min_asv_cauchy}) matches the value from the simulation. 

\begin{figure}[tb]
\begin{minipage}[b]{1.0\linewidth}
  \centering
  \centerline{\epsfig{figure=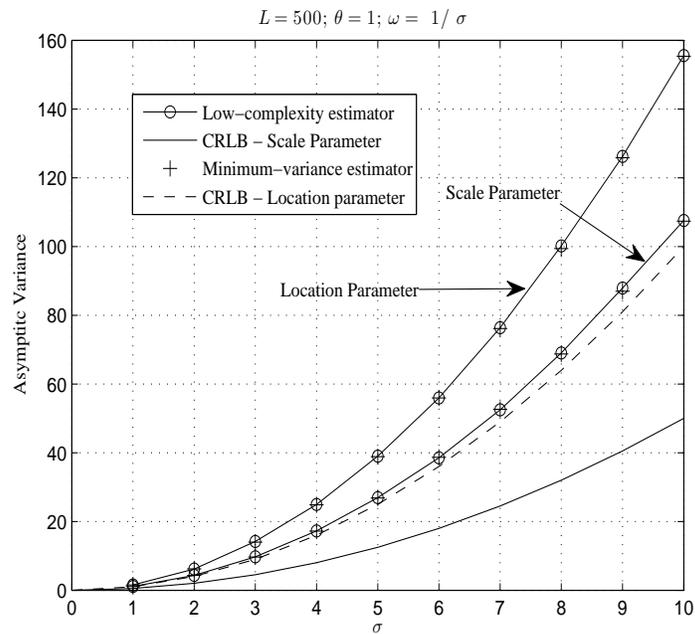,width=10.5cm,height=9cm}}
\end{minipage}
\caption{Asymptotic variance and CRLB vs. scale parameter. Sensing noise is Laplace distributed.} \label{Fig:laplace}
\end{figure}

In the per-sensor power case, when the sensing noise is Gaussian distributed, it has been argued in Section \ref{sssec:gauss_pspc} that the estimators are optimal if $\omega\to0$, while the sensed data are not sent to the sensors if $\omega = 0$. This discontinuity is demonstrated in Figure \ref{Fig:gauss_pspc_diff_L}. It can be seen that for small $L$, as $\omega\to0$, the performance suffers. However, for a larger number of sensors, the asymptotic variance is minimized for smaller $\omega$. 

In Figure \ref{Fig:gauss}, ${\rm AsV}_{\hat{\lp}} (\omega)$ and ${\rm AsV}_{\hat{\scpar}} (\omega)$ are plotted versus $\scpar$ when the sensing noise is Gaussian for the per-sensor power constraint case. The value of $\omega = 0.01$ is used. When compared against the CRLB, in the case of Gaussian sensing noise, both the estimators are asymptotically efficient, since the asymptotic variances are the same as the respective values of the CRLB. 

\begin{figure}[tb]
\begin{minipage}[b]{1.0\linewidth}
  \centering
  \centerline{\epsfig{figure=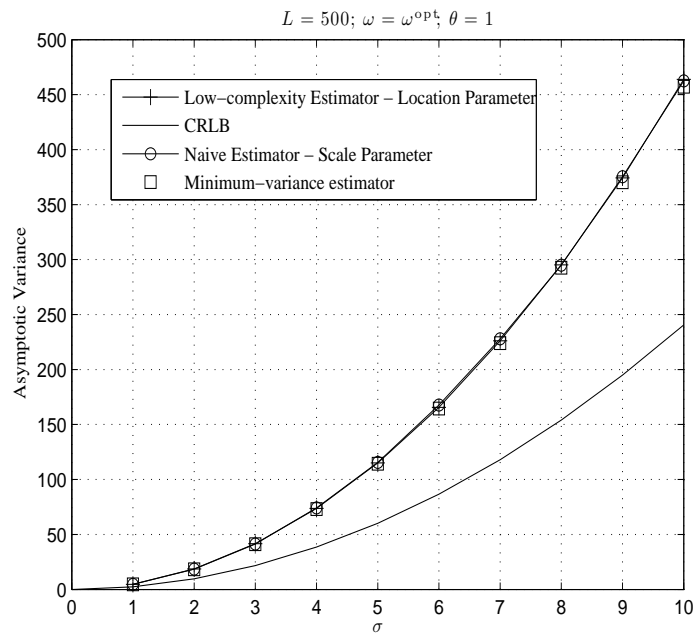,width=10.5cm,height=9cm}}
\end{minipage}
\caption{Performance vs. $\scpar$. Sensing noise is Cauchy distributed.} \label{Fig:cauchy}
\end{figure}

In Figure \ref{Fig:laplace}, the sensing noise is Laplace distributed. The performance of the estimators of $\lp$, $\scpar$ and $\gamma$ are plotted against $\scpar$ for the per-sensor constraint case. For transmissions from these sensors, $\omega = \omega_{\theta}^{*} = 1/\scpar$. In this case, the estimators are not asymptotically efficient as the asymptotic variances are larger than the CRLB. 

Cauchy distributed sensing noise was considered for the results shown in Figure \ref{Fig:cauchy}. The estimators of both the location parameter and the scale parameter are plotted versus $\scpar$ for the per-sensor power constraint case. Here, $\omega = \omega^{*} = 0.7968/\scpar$. Both estimators have the same performance. Additionally, both parameters have the same CRLB, which are lower than the asymptotic variances of the location parameter and the scale parameter. Therefore, in the case when the sensing noise is Cauchy distributed, the estimators are also not asymptotically efficient. 

\section{Conclusions}
\label{sec:conclusions}

A problem of simultaneous distributed estimation of the scale parameter and location parameter of a signal embedded in noise was considered for arbitrary sensing noise distributions. Sensors observe a parameter in sensing noise and transmit the observations using a constant-modulus phase modulation scheme. The sensors transmit the observations over a Gaussian multiple-access channel to a fusion center. Due to the additive nature of the channel, the signal received at the FC converges to the characteristic function of the sensing noise distribution as the number of sensors grows large. The phase-modulation scheme presented is robust to impulsive sensing noise distributions and ensures a fixed transmit power irrespective of the signal or noise realization. Two cases of sensor power are considered, one with a total power constraint across all the sensors, and another imposing a per-sensor power constraint. 

At the fusion center, an estimation scheme is presented that is used to estimate the mean, variance, and SNR of the observations, by using only a single set of transmissions from the sensors. Estimators are used to estimate a location parameter, $\lp$, and a scale parameter, $\scpar$. 
For each case of sensing noise distribution, the optimum transmission parameter, $\omega$, was calculated. 

The estimates of the scale parameter and the location parameter were combined to construct an estimator for the SNR of the observations. It was shown that this was an asymptotically minimum-variance estimator. The performance of the SNR estimator was also evaluated. 

For each estimator, the value of $\omega$ that minimized the asymptotic variance was determined. It was shown that when the asymptotic variances of the location parameter and the scale parameter were quasi-convex in $\omega$, the optimum value of $\omega$ for the SNR estimator lies between the optimum values of $\omega$ for the location parameter and the scale parameter. 

In the case of per-sensor power constraint, the asymptotic efficiency of the estimators was also evaluated. When the sensing noise distribution is Gaussian distributed, the estimators are shown to be asymptotically efficient. For the other sensing noise distributions considered in the paper, the estimators are not asymptotically efficient, though they may have a better asymptotic variance than in the case of the Gaussian sensing noise. 

\appendices

\section{Proof of Theorem \ref{thm:compare_estimators}}
\label{app:proof_thm_compare_estimators}

Using the values of $\mathbf{z}_{L}$ and $\bar{\mathbf{z}}$ and using $\boldsymbol{\Sigma}$ from (\ref{eqn:sigma_defn}), the estimator in (\ref{eqn:est_lp_scpar}) can be simplified to
\begin{align}
\left[ \hat{\lp}^{\rm opt} \quad \hat{\scpar}^{\rm opt} \right]^{T} = \argmin_{\lp, \scpar} & \Big\{|z_{L}|^{2} \left[ \varphi_{\eta} (2 \scpar \omega) + 1 \right] \left[ 1 + P \varphi_{\eta}^{2} (\scpar \omega) \right] \nonumber \\ 
& + P \varphi_{\eta}^{2} (\scpar \omega) \left[ 1 - (z_{L}^{R})^{2} \cos^{2} (\omega \lp) -(z_{L}^{I})^{2} \sin^{2} (\omega \lp) - z_{L}^{R} z_{L}^{I} \sin (2 \omega \lp) \right] \nonumber \\
& + P \varphi_{\eta}^{2} (\scpar \omega) \left[ \varphi_{\eta} (2 \scpar \omega) - \varphi_{\eta}^{2} (\scpar \omega) \right] \nonumber \\
& - P \varphi_{\eta}^{2} (\scpar \omega) \left[ z_{L}^{R} \cos (\omega \lp) + z_{L}^{I} \sin (\omega \lp) \right]^{2} \left[ \varphi_{\eta} (2 \scpar \omega) - \varphi_{\eta}^{2} (\scpar \omega) \right] \nonumber \\
& - 2 \sqrt{P} \varphi_{\eta} (\scpar \omega) \left[ z_{L}^{R} \cos (\omega \lp) + z_{L}^{I} \sin (\omega \lp) \right] \left[ 1 + \varphi_{\eta} (2 \scpar \omega) \right] \Big\}.
\label{eqn:min_prbm2}
\end{align}
Defining $s(\lp) \mathop{:=} z^{R} \cos (\omega \lp) + z^{I} \sin (\omega \lp)$, the problem is rewritten first as
\begin{align}
\left[ \hat{\lp}^{\rm opt} \quad\hat{\scpar}^{\rm opt} \right]^{T} = \argmin_{s(\lp)} & \Big\{ |z_{L}|^{2} \left[ \varphi_{\eta} (2 \scpar \omega) + 1 \right] \left[ 1 + P \varphi_{\eta}^{2} (\scpar \omega) \right] \nonumber \\
& + P \varphi_{\eta}^{2} (\scpar \omega) \left[ 1 - s^{2} (\lp) \right] \left[ 1 + \varphi_{\eta} (2 \scpar \omega) \right] \nonumber \\
& - 2 \sqrt{P} \varphi_{\eta} (\scpar \omega) s (\lp) \left[ 1 + \varphi_{\eta} (2 \scpar \omega) \right] \Big\}.
\label{eqn:min_prbm4}
\end{align}
Examining the first derivative of the objective function in (\ref{eqn:min_prbm4}), we have that $\lp$ solves (\ref{eqn:min_prbm4}) if and only if it solves
\begin{equation}
s(\lp) = |z_{L}|. \label{eqn:min_lp_naive}
\end{equation}
The minimization problem in (\ref{eqn:min_prbm2}) can now be reformulated as
\begin{equation}
\hat{\scpar}^{\rm opt} = \argmin_{\scpar > 0} \left[ 1 + \varphi_{\eta}(2 \scpar \omega) \right] \left[ |z_{L}| - \sqrt{P} \varphi_{\eta}(\scpar \omega) \right]^{2}.
\label{eqn:min_prbm3}
\end{equation}
For large $L$, it can be seen from (\ref{eqn:zl_defn}) that the effect of the channel noise is diminished, and with high probability, $|z_{L}| \leq \sqrt{P}$. The objective function is minimized when $|z_{L}| - \sqrt{P} \varphi_{\eta}(\scpar \omega) = 0$, which is identical to (\ref{eqn:z_abs}). Substituting $|z_{L}| = \sqrt{P} \varphi_{\eta}(\scpar \omega)$ in (\ref{eqn:min_lp_naive}), the equation in (\ref{eqn:z_arg}) is obtained, completing the proof.

\section{Proof of Theorem \ref{thm:opt_omega}}
\label{app:proof_thm_opt_omega}

With $\text{AsV}_{\hat{\gamma}} (\omega) = \alpha \text{AsV}_{\hat{\lp}} (\omega) + \beta \text{AsV}_{\hat{\scpar}} (\omega)$, $\alpha >0$ and $\beta>0$, it is required to prove Theorem \ref{thm:opt_omega}. Three cases are considered:

\subsubsection{Both $\text{AsV}_{\hat{\lp}} (\omega)$ and $\text{AsV}_{\hat{\scpar}} (\omega)$ satisfy condition (c1)}

When $\text{AsV}_{\hat{\lp}} (\omega)$ and $\text{AsV}_{\hat{\scpar}} (\omega)$ are monotonically non-decreasing, both will have their infima as $\omega\to0$. $\text{AsV}_{\hat{\gamma}} (\omega)$ will also be monotonically non-decreasing and will have its infimum as $\omega \to 0$. Similarly, when both $\text{AsV}_{\hat{\lp}} (\omega)$ and $\text{AsV}_{\hat{\scpar}} (\omega)$ are monotonically non-increasing, the minima of $\text{AsV}_{\hat{\lp}} (\omega)$, $\text{AsV}_{\hat{\scpar}} (\omega)$ and $\text{AsV}_{\hat{\gamma}} (\omega)$ will lie at $\omega = 2\pi/\lp_{R}$. Finally, when one of the two functions is monotonically non-increasing, and the other is monotonically non-decreasing, $\omega_{\lp}^{*}$ and $\omega_{\scpar}^{*}$ will be $0$ or $2\pi/\lp_{R}$. Since $\omega_{\gamma}^{*}$ can lie only in $[0,2\pi/\lp_{R}]$, the proof is complete for this case. 

\subsubsection{Both $\text{AsV}_{\hat{\lp}} (\omega)$ and $\text{AsV}_{\hat{\scpar}} (\omega)$ satisfy condition (c2)}

If $\omega_{\lp}^{*}$ minimizes $\text{AsV}_{\hat{\lp}} (\omega)$, $AsV_{\hat{\lp}}^{\prime} (\omega_{\lp}^{*}) = 0$, where $AsV_{\hat{\lp}}^{\prime} (\omega_{\lp}^{*})$ is the first derivative of $\text{AsV}_{\hat{\lp}} (\omega)$ with respect to $\omega$, evaluated at $\omega_{\lp}^{*}$. Similarly if $\text{AsV}_{\hat{\scpar}} (\omega)$ is minimized at $\omega_{\scpar}^{*}$, $AsV_{\hat{\scpar}}^{\prime} (\omega_{\scpar}^{*}) = 0$. 

From (\ref{eqn:SNR_asv}), the expression for the asymptotic variance of $\hat{\gamma}$ is given by 
\begin{equation}
\text{AsV}_{\hat{\gamma}} (\omega) = \alpha  \text{AsV}_{\hat{\lp}} (\omega) + \beta  \text{AsV}_{\hat{\scpar}} (\omega),
\label{eqn:asv_gamma_expn}
\end{equation}
where $\alpha = 4 \lp^{2}/\scpar^{4} > 0$ and $\beta = 16 \lp^{2}/\scpar^{6} >0$. If $\omega_{\gamma}^{*}$ is the minimizer of $\text{AsV}_{\hat{\gamma}} (\omega)$, it is required to verify that
\begin{equation}
AsV_{\hat{\gamma}}^{\prime} (\omega_{\gamma}^{*}) = 0. 
\label{eqn:min_asv_gamma}
\end{equation}
The left-hand side of (\ref{eqn:min_asv_gamma}) can be rewritten using (\ref{eqn:asv_gamma_expn}) so that the condition for $\omega_{\gamma}^{*}$ to be the minimizer of $\text{AsV}_{\hat{\gamma}} (\omega)$ is given by
\begin{equation}
\frac{ \text{AsV}_{\hat{\lp}}^{\prime} (\omega_{\gamma}^{*})} { \text{AsV}_{\hat{\scpar}}^{\prime} (\omega_{\gamma}^{*})} = -\frac{\beta}{\alpha}.
\label{eqn:asv_gamma_diff_cond}
\end{equation}
The right hand side of (\ref{eqn:asv_gamma_diff_cond}) is negative. This happens only when one of the slopes of $\text{AsV}_{\hat{\lp}} (\omega)$ and $\text{AsV}_{\hat{\scpar}} (\omega)$ is positive and the other is negative. By using the quasi-convexity of $\text{AsV}_{\hat{\lp}} (\omega)$ and $\text{AsV}_{\hat{\scpar}} (\omega)$, it can be seen that when the functions are quasi-convex, and when $\omega_{\lp}^{*}<\omega_{\scpar}^{*}$, the $\omega$ axis can be divided into three regions: (i) $\omega < \omega_{\lp}^{*}$, where both $\text{AsV}_{\hat{\lp}} (\omega)$ and $\text{AsV}_{\hat{\scpar}} (\omega)$ have negative slope; (ii) $\omega_{\lp}^{*} < \omega < \omega_{\scpar}^{*}$, where $\text{AsV}_{\hat{\lp}} (\omega)$ has a positive slope and $\text{AsV}_{\hat{\scpar}} (\omega)$ has a negative slope; and (iii) $\omega > \omega_{\scpar}^{*}$, where $\text{AsV}_{\hat{\lp}} (\omega)$ and $\text{AsV}_{\hat{\scpar}} (\omega)$ both have positive slope \cite[pp. 99]{boyd2004}. Therefore, the condition in (\ref{eqn:asv_gamma_diff_cond}) is satisfied only when $\omega_{\lp}^{*} \leq \omega_{\gamma}^{*} \leq \omega_{\scpar}^{*}$. A similar argument can be made when $\omega_{\lp}^{*}>\omega_{\scpar}^{*}$. 

\subsubsection{One of $\text{AsV}_{\hat{\lp}} (\omega)$ and $\text{AsV}_{\hat{\scpar}} (\omega)$ satisfies condition (c1) and the other satisfies condition (c2)}

Let $\text{AsV}_{\hat{\lp}} (\omega)$ satisfy condition (c1) and let $\text{AsV}_{\hat{\scpar}} (\omega)$ satisfy condition (c2). The condition in (\ref{eqn:asv_gamma_diff_cond}) will need to hold in order to prove Theorem \ref{thm:opt_omega}. If $\text{AsV}_{\hat{\lp}} (\omega)$ is monotonically non-decreasing, $\omega_{\lp}^{*} \to 0$, and (\ref{eqn:asv_gamma_diff_cond}) is satisfied when $\omega_{\gamma}^{*} \leq \omega_{\scpar}^{*}$. When $\text{AsV}_{\hat{\lp}} (\omega)$ is monotonically non-increasing, $\omega_{\lp}^{*} = 2\pi/\lp_{R}$, and (\ref{eqn:asv_gamma_diff_cond}) is satisfied when $\omega_{\gamma}^{*} \geq \omega_{\scpar}^{*}$. A similar argument can be made when $\text{AsV}_{\hat{\lp}} (\omega)$ satisfies condition (c2) and $\text{AsV}_{\hat{\scpar}} (\omega)$ satisfies condition (c1), completing the proof. 

\bibliography{phase_est}
\end{document}